\documentclass[11pt]{article}
\usepackage{epsfig}
\begin{document}
\vspace*{1.1cm}
\begin{center}
{\LARGE \bf The Apparent Anomalous, Weak, Long-Range \\ Acceleration  
of Pioneer 10 and 11$^{\dag\S}$}
\end{center}
\begin{center}
{Slava G. Turyshev,$^a$ John D. Anderson,$^b$ Philip A. 
Laing,$^c$ \\ Eunice L.Lau,$^d$  Anthony S. Liu,$^e$ and Michael Martin
Nieto$^f$ }  
\end{center}

\begin{center}
{\vskip 10pt
\it $^{a,b,d}$ Jet Propulsion Laboratory, California Institute
of  Technology, Pasadena, CA 91109 \\ \vskip 10pt 
$^c$ The Aerospace Corporation, 2350 E. El Segundo Blvd., 
El Segundo,  CA 90245-4691  \\ \vskip 10pt
$^e$ Astrodynamic Sciences, 2393 Silver Ridge Ave., Los Angeles, 
CA 90039  \\ \vskip 10pt
$^f$ Theoretical Division (MS-B285), Los Alamos National Laboratory,\\
University of California, Los Alamos, NM 87545}
\end{center} 

\baselineskip=.175in
 
\begin{abstract}  
Recently  we reported that radio Doppler data generated by NASA's Deep Space
Network (DSN) from the Pioneer 10 and 11 spacecraft indicate an
apparent anomalous, constant, spacecraft acceleration
with a magnitude  $\sim 8.5\times 10^{-8}$ cm s$^{-2}$, directed towards the
Sun \cite{anderson}. Analysis of similar Doppler and ranging data from the
Galileo and Ulysses spacecraft yielded ambiguous results for the anomalous
acceleration, but it was useful in that it ruled out the
possibility of a systematic error in the DSN Doppler system that could
easily have been mistaken as a spacecraft acceleration.  Here we present
some new results, including a critique suggestions that the anomalous
acceleration could be caused by collimated thermal emission. Based
partially on a further data for the Pioneer 10 orbit determination,
the data now spans January 1987 to July 1998, our best estimate of the
average Pioneer 10 acceleration directed towards the Sun is $\sim 7.5
\times 10^{-8}$ cm s$^{-2}$.  
\end{abstract}  
 
\section{Introduction}
\label{sec:intro}

Detailed analyses of radio metric data from   distant
spacecraft in the solar system have revealed an anomalous acceleration
acting on  Pioneer 10 and 11, with supporting data from  Galileo, and
Ulysses spacecraft.  These data indicated existence of an apparent
anomalous, constant, acceleration acting on the   spacecraft with a
magnitude  $\sim 8.5\times 10^{-8}$ cm/s$^2$, directed towards the
Sun \cite{anderson}.  Two independent codes and physical strategies were
used to  analyze the data.   A number of potential causes have been ruled
out.  In this paper we  report on  further progress in this study.

We concentrate on the analysis of the Pioneer 10 and 
11 spacecraft Doppler data. We will discuss  scenarios that
involve excess power and heat generated  by the Radioisotope
Thermoelectric Generators (RTGs).  We will present our estimates for the
corresponding effects in order to demonstrate that these
mechanisms  can not as yet  explain  the reported effect.
\vfill
\noindent\underline{\hskip 6.5cm}
\vskip -5pt
{\footnotesize $^\dag$ Presented at XXXIV-th Rencontres de 
Moriond Meeting on Gravitational Waves and Experimental Gravity. 
Les Arcs, Savoi, France  (January 23-30, 1999) }

\section{Study of the anomalous acceleration}

The Pioneer spacecraft are excellent for dynamical astronomy studies. 
Due to their spin-stabilization and their great distances, 
a minimum number of Earth-attitude reorientation maneuvers are required. 
To obtain the S-band Doppler data from the Pioneer spacecraft, NASA used
the Jet Propulsion Laboratory's (JPL) Deep Space Network (DSN).  
The signals were actively reflected by a transponder on the spacecraft and
calculation of the motions of the spacecraft were made based on the
resulting Doppler shift in the signals.  This data was used to
determine the Pioneers position, velocity and  the magnitudes of the
orientation maneuvers. 

\subsection{The studies of unmodeled acceleration at JPL.}
\label{subsec:accel}

Beginning in 1980, when at 20 AU the solar radiation pressure 
acceleration had decreased to  
$< 5 \times 10^{-8}$ cm/s$^2$,  JPL's 
Orbit Determination Program (ODP) analysis of unmodeled 
accelerations (at first with the faster-moving Pioneer 10) 
found that the biggest systematic error in the acceleration residuals
is a constant bias of $a_P \sim (8\pm 3) \times 10^{-8}$ cm/s$^2$, 
directed {\it toward} the Sun 
(to within the accuracy of the Pioneers' antennae).  


We ultimately concluded \cite{anderson}   
that there is an unmodeled  acceleration, $a_P$, towards the Sun  of
$(8.09\pm0.20)\times 10^{-8}$  cm/s$^2$ for Pioneer 10 and of $(8.56\pm
0.15) \times 10^{-8}$ cm/s$^2$   for Pioneer 11.  The error is determined
by use of a  five-day batch sequential filter with radial acceleration as
a stochastic parameter subject to white Gaussian noise ($\sim$ 500
independent five-day samples of radial acceleration).
No magnitude variation of $a_P$ with distance was found, within a  
sensitivity of 2 $\times$ 10$^{-8}$ cm/s$^2$ over a 
range of 40 to 60 AU. 
All errors are from the covariance matrices associated with the
least--squares   analysis.
The assumed data errors are larger than the standard error on the 
post--fit residuals \cite{anderson}.

The  observed effect may be expressed by the following simple expression:
\begin{eqnarray}
{\nu}_{obs}&=&\nu_{model}\times 
\left[\,1 - \frac{a_P \cdot t}{c}\,\,\right], 
\end{eqnarray}
\noindent where ${\nu}_{obs}$ is the frequency of the re-transmitted
signal observed by a DSN antennae, while $\nu_{model}$ is the
predicted frequency of that signal.  Our analyses were were modeled to
include the effects of planetary perturbations, radiation pressure, the
interplanetary  media, general relativity, and  bias and drift in the
Doppler signal.  Planetary coordinates and the solar system masses  were
obtained using JPL's Export Planetary  Ephemeris DE200.  The analyses
calculated Earth's polar motion and its  non-uniform rotation using the
International Earth  Rotation Service.
 
The models   account for a number of
post-New\-tonian perturbations in the dynamics of the planets, the Moon,
and spacecraft: i)  models for light propagation  are correct to order
$(v/c)^2$,  ii)  the equations of motion of extended celestial bodies are
valid to order  $(v/c)^4$. Non-gravitational effects, such as solar
radiation pressure and precessional attitude-control maneuvers, make
small contributions to the apparent acceleration we have observed.   The
solar radiation pressure decreases as $r^{-2}$.   As 
previously indicated for the Pioneers, 
at distances $>$10-15 AU it produces an  acceleration that is 
much less than $8\times 10^{-8}$ cm/s$^2$, directed 
{\it away} from the Sun. (The solar wind is
roughly a factor of 100 smaller than this.) 

 As possible ``perturbative forces" to explain this bias, we considered
gravity  from the Kuiper belt, gravity from the galaxy,
spacecraft ``gas leaks,'' errors in the planetary ephemeris, and
errors in the accepted values of the
Earth's orientation, precession,  and nutation. 
None of these ``forces" could  explain the apparent acceleration, 
and some were two orders of magnitude or more too small.

\subsection{An error in the code? --- The Aerospace Corporation's result.}
\label{subsec:aero}
With no explanation of this data in hand, our attention focused on
the possibility that there was some error in  JPL's ODP.  To investigate
this, an independent analysis of the raw data using The
Aerospace Corporation's Compact High Accuracy Satellite Motion Program
(CHASMP), which was developed independently
of JPL's ODP, was performed. Although, by necessity, both programs use
the same physical principles,  planetary ephemeris, and timing and polar
motion inputs, the algorithms are  otherwise quite different. If there
were an error in either program, they would not agree. (Common program
elements continue to be investigated.)

The CHASMP analysis of Pioneer 10 data
also showed an unmodeled acceleration 
in a direction along the radial toward the  Sun. 
The value is $(8.65 \pm 0.03) \times 10^{-8}$ cm/s$^{2}$, 
agreeing with JPL's result. 
The smaller error here is because the CHASMP analysis 
used a batch least-squares fit over the whole orbit, not looking 
for a variation of the magnitude of $a_P$ with distance.

Without using the apparent acceleration, 
CHASMP shows a steady frequency 
drift\footnote{The JPL and DSN convention for Doppler frequency shift is
$\Delta \nu = \nu - \nu_0$, where $\nu$ is the measured frequency and
$\nu_0$ is the reference frequency. It is positive for a spacecraft
receding from the tracking station (red shift), and negative for a
spacecraft approaching the station (blue shift), just the opposite of the
usual convention.} of about
$-6 \times 10^{-9}$ Hz/s, or 1.5 Hz over 8 years 
(one-way only). 
The drift in the Doppler residuals (observed minus computed data) is 
clear, definite, and  cannot be removed without   the added
acceleration, $a_P$. 
If there were a systematic drift in the atomic clocks of the DSN  or
in the time-reference standard signals, this would appear   like a
non-uniformity of time; i.e., all clocks would be changing with a 
constant acceleration. We now have been able to rule out this
possibility. 

In addition to our previous analysis \cite{anderson},   
we have examined numerous ``time'' models (in conjunction with our
studies of Galileo and Ulysses spacecraft radio metric data), searching 
for any (possibly radical) solution, namely:  i). {\it Drifting Clocks.}
This model adds a constant  acceleration term to the Station Time (ST) 
clocks; i.e.. in the ST-UTC (Universal Time Coordinates)  time
transformation. The model fit Doppler well for Pioneer 10, Galileo,  and
Ulysses but failed to model range data for  Galileo and Ulysses.    
ii).  {\it Quadratic Time Augmentation.} 
This model adds a quadratic-in-time augmentation
to the IAT-ET (International Atomic Time-Ephemeris Time) 
time transformation. 
The model fits Doppler fairly well but range very badly.  
iii).   {\it Frequency Drift.}  
This model adds a constant frequency drift to the 
reference frequency. The model also fits Doppler well but again fits 
range poorly. 
iv). {\it Expanding Space}. This model adds a quadratic in time 
term to the 
light time, thus mimicking a line of sight acceleration of the spacecraft.
The model fits both Doppler and range very well but the coefficient
of the quadratic is negative for Pioneer 10 and Galileo while positive for
Ulysses.  
v) {\it Speed of Gravity}. This model adds a ``light time" delay  
to the actions of the Sun and planets upon the spacecraft. The model fits
Pioneer 10 and Ulysses well.  But the Earth flyby of Galileo  fit was
terrible, with Doppler  residuals as high as 20 Hz.  

All these models were rejected due
either to poor fits or to inconsistent  solutions among spacecraft. 

\section{Influence of the excess power and heat from RTGs.}
\label{sec:syst}

One might argue that a possible systematic explanation of the residuals
is  non-isotropic thermal radiation.   Pu$^{238}$ radioactive thermal
generators (RTGs) power the Pioneers. 
 
\subsection{The heat coming from the RTGs.}
\label{subsec:katz}

Let us discuss the anisotropic heat reflection off of the back  
of the spacecraft high-gain antennae, the heat coming from the
RTGs \cite{katz}. Before launch the four RTGs delivered a total electrical
power  of 160 W (now $\sim$ 70-75 W), 
from  a total thermal fuel inventory of 2580 W (now $\sim$ 2090 W).   
Presently  $\sim 2000$ W of RTG heat must be dissipated.  
Only $\sim(70-85)$ W  of directed power could explain the 
anomaly. Therefore, in principle there is enough 
power to explain the anomaly this way.  However, 
there are two reasons that preclude such a mechanism, namely:

i). {\sf The spacecraft geometry.}  The RTGs are located at the end
of booms, and  rotate about the craft in a plane that contains the
approximate base of  the antenna.  From the RTGs the antenna is thus seen
``edge on" and   subtends a solid angle of $\sim$ 1.5 \% 
of $4\pi$ steradians.   This already means  
the proposal could provide at most $\sim 30$ W.  But there 
is more. 

ii) {\sf The RTG's radiation pattern.}  
The above estimate was based on the assumption that  
the RTGs are spherical black bodies.  But they  are not.  
The main bodies of the RTGs are cylinders and they 
are grouped in two packages of two.  Each 
package has the two cylinders end to end extending away from 
the antenna. 
Every RTG has six fins that go radially out from the cylinder.  
Thus, the fins are ``edge on" to the antenna (the fins point 
perpendicular to the cylinder axes).  Ignoring edge 
effects, this means that only 2.5 \% of the surface area of the RTGs  
is facing the antenna.  Further, for better radiation from the fins,  
the Pioneer SNAP 19 RTGs had larger fins than the earlier test models, and 
the packages were insulated so that the end caps had lower temperatures 
and radiated less than the cylinder/fins \cite{tele}.
As a result of such a design, the vast majority of the 
heat radiated by the RTG's is symmetrically directed 
in space unobscured by the antenna

We conclude that this mechanism does not provide enough power to  
explain the Pioneer anomaly.

\subsection{Non-isotropic radiative cooling of the spacecraft.}
\label{subsec:mainbus}

There is also the possibility that the 
anomalous acceleration seen in the Pioneer 10/11
spacecraft can be, ``explained, at least in part, by
non-isotropic  radiative cooling of the spacecraft "\cite{murphy}.  So, the
question is, does  ``at least in part" mean this effect  comes  near to
explaining the anomaly?  We argue it does not. 

Consider radiation of the power of the main-bus electrical 
systems  from  the rear of the craft.   
For the Pioneers, the aft has a louver system, and 
``the louver system acts to control the heat 
rejection of the radiating platform... A bimetallic spring, thermally
coupled radiatively to the platform, provides the motive force for 
altering the angle of each blade. In a closed position the heat rejection 
of the platform is minimized by virtue of the ``blockage'' of the 
blades while open louvers provide the platform with a nearly 
unobsructed view of space'' \cite{piodoc}.

If these louvers were open,   this mechanism could produce 
a comparable effect.  However, by 9 AU the actuator spring 
temperature had already reached $\sim$40$^{\circ}$F. This means the 
louver doors were   closed  (i.e.,  the louver angle was zero) 
from there on out. Thus, from our quoting of the radiation properties
above,  the contribution of the thermal radiation to the Pioneer anomalous
acceleration should be {\em negligibly} small.   After the above
time, we reach our data region.  In 1984 Pioneer  10 was at about 33 AU
and the power was about 105 W.  (Always reduce  the total power numbers
by 8 W to account for the radio beam power.) In (1987, 1992, 1996) the craft
was at $\sim$(41, 55, 65) AU and the power was $\sim$(95, 80, 70) W. 
The louvers were inactive, and no decrease in $a_P$ was 
seen\footnote{Any change of the louver angle should result
in a spin change due to the thermal radiation. This is because  
of the orientation of the lovers around the bus on the spacecraft.
We detect no such a change.}.

We conclude that this proposal can not explain the anomalous Pioneer 
acceleration.  

\subsection{Could the Helium pressure produced  within
the RTGs  be  the cause for the  acceleration?}
\label{subsec:helium}

Another possible systematic from on-board the spacecraft,
is from the He  build-up in the RTGs due to the $\alpha$-decay of
Pu$^{238}$.  Is there any way that this permeating He could be causing 
$a_P$?

To make this mechanism work, one would need that the He leackage from
the RTGs  be  preferentially directed away from the Sun,  
with a velocity large enough to cause the acceleration. 
The SNAP-19 Pioneer RTGs were designed in a such a way that the He
pressure has not been totally  contained within the Pioneer heat source 
over the life of RTGs. Instead, the Pioneer heat source
contains a pressure relief device which allows the generated He to vent
out of the heat source and into the thermoelectric converter.  (The
strength member and the capsule clad contain small holes to permit He to
escape into the thermoelectric converter.) The thermoelectric converter
housing-to-power output receptacle interface is sealed with a viton
O-ring. This means that, due to permeation, the  gas within the converter
is expected to be released to the space environment throughout the mission
life of the Pioneer RTGs. 

From the size of the fuel pucks, the total volume of fuel is about 904
cm$^3$.  The fuel is   PMC Pu conglomerate.   
The amount of Pu$^{238}$ in this fuel is about 5.8 kg. With a decay
constant of 87.74 years, that means the rate of He production (from Pu
decay) is about 0.8 gm/year, assuming it all leaves the cermet. 
Finally, $kT$ on the surface of the  RTGs corresponds to less than
$10^{3}$ km/sec for He. (When looking for gas leaks, see below, we found
that 2 gr/year mass of hydrazine  could produce $a_P$ {\em if  it
came out} at nozzle speed  of about 3.4 km/sec, {\em all directed}.)

So, we can rule out this helium permeating through the O-rings
as the cause of our effect. 


\section{Recent  results}
\label{sec:new}
\begin{figure}[ht]
 \begin{center}
\noindent    
\psfig{figure=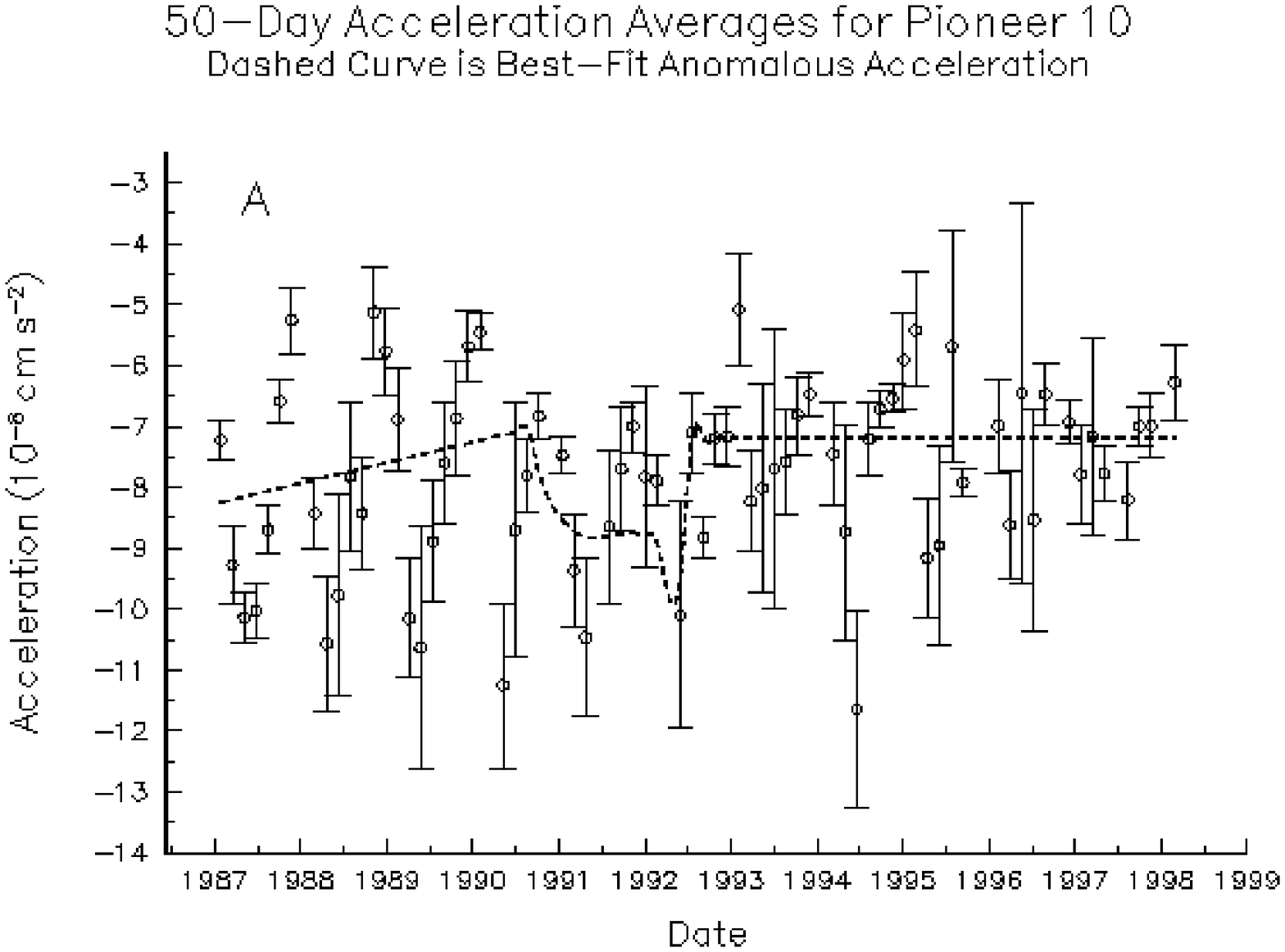,width=125mm,height=90mm}
    \psfig{figure=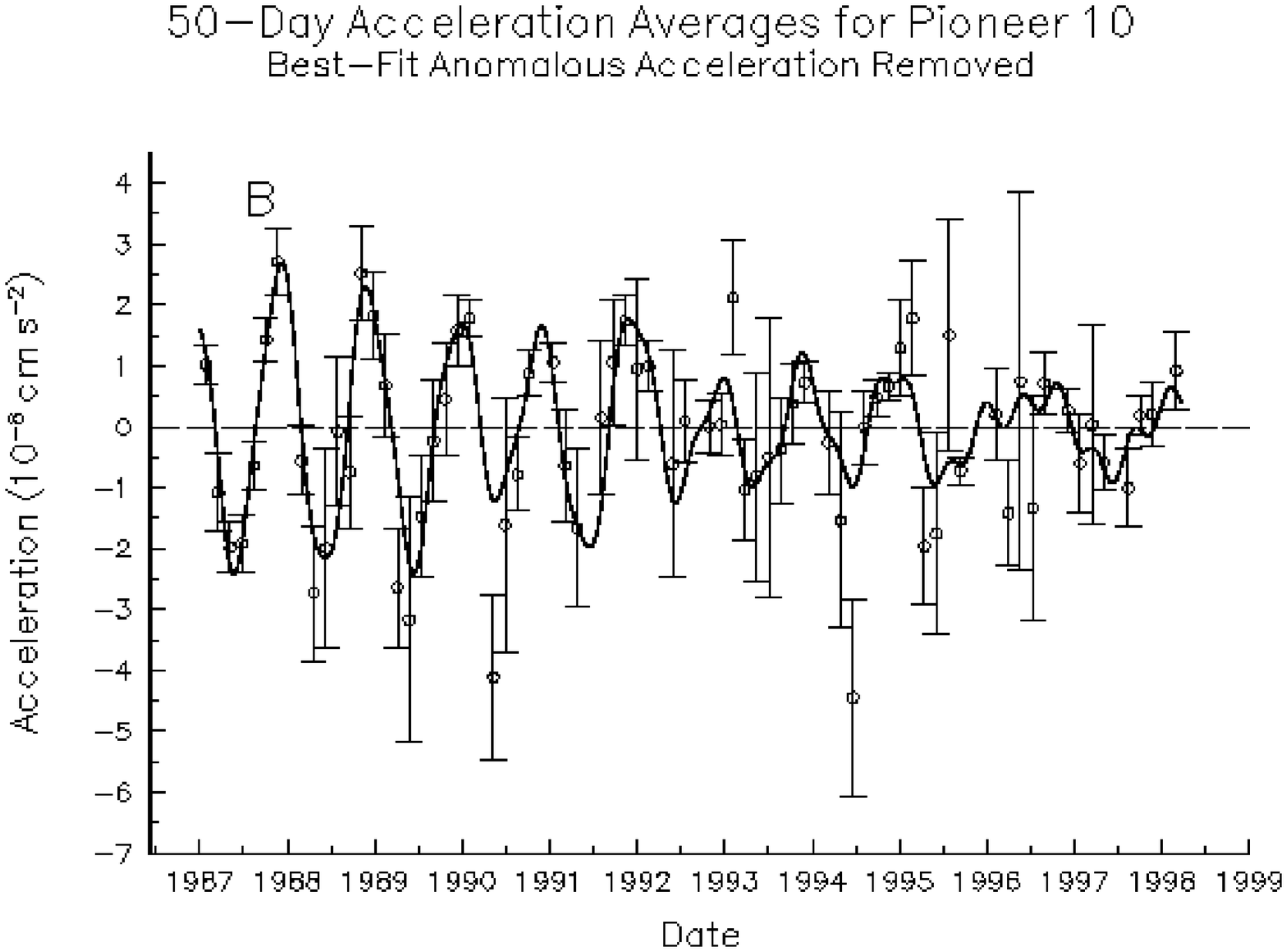,width=125mm,height=90mm}
    \caption{The Pioneer 10 Doppler data: 50-day averages of anomalous 
              acceleration -- January 1987 to July 1998. 
      \label{fig:data}}
 \end{center}
\end{figure}

Recently  we began using new JPL software (SIGMA) to reduce the Pioneer 10
Doppler data to 50-day  averages of acceleration, extending from January
1987 to July 1998,  over a distance interval from 40 to 69 AU
(see Figure \ref{fig:data}).   
Before mid-1990, the spacecraft rotation rate changed (slowed) by  
about -0.065 rev/day/day. Between mid-1990 and mid-1992 the 
spin-deceleration increased to -0.4 rev/day/day. But after mid-1992 the 
spin rate remained $\sim$ constant.   
In units of 10$^{-8}$ cm/s$^{2}$, the mean acceleration levels obtained 
by SIGMA from the Doppler data in these periods are:
$(7.94 \pm 0.11)$ before mid-1990, $(8.39 \pm 0.14)$ between mid-1990 
and mid-1992, and $(7.29 \pm 0.17)$ after mid-1992. 
[Similar values $(8.27\pm 0.05, ~8.77\pm 0.04, ~7.76\pm 0.08)$ 
were obtained using CHASMP.] 
We detect no long-term deceleration changes from mid-1992 to mid-1998, 
and only two spin-related discontinuities over the entire data period.
\noindent 

We have also performed a number of tests of the internal consistency of
our analysis of $a_P$.  Internal tests included numerous examinations of
the fit results for various  aspects of the theoretical model related to
station coordinates, Earth orientation parameters, precession, nutation, 
instrumental clock stability, interplanetary plasma effects, and
relativistic effects. The numerical stability of the estimation algorithm
and its  computer implementation are also considered. Finally, the
stability of the  solution was examined in detail, in terms of
time-dependent changes of  both the
$a_P$ and the spacecraft spin-down rate.

The internal consistency tests indicate that, in addition to the formal 
uncertainties, there is evidence for a systematic mismodeling which
results in an annual periodic term (plot {\tt B} in Figure \ref{fig:data}).
This term has been found in the residuals of the both Pioneers and
is currently being investigated. Such systematic errors lead to estimates
of realistic uncertainties that are approximately two times the formal 
uncertainties. 

Assume that the slowing of the spin rate was caused by
spacecraft  systems that also account for a few \% systematic effect.  
Then, excluding other biases, such as the radio beam increasing the
anomaly,  we should adopt the post-1992 value as the most accurate
measure of  the anomalous acceleration of Pioneer 10.

As stated before, we believe the most plausible
explanation of the anomaly is systematics, such as  radiant heat 
or gas leaks.    But no such explanation has yet been
demonstrated. Clearly, more analysis, 
observation, and theoretical work are called for.  
Further details will appear elsewhere.   

\section*{Acknowledgments}

This work was supported by the Pioneer Project, NASA/Ames Research
Center, and was performed at the Jet Propulsion Laboratory, California
Institute of Technology, under contract with NASA. P.A.L. and A.S.L.
acknowledge support by a grant from NASA through the Ultraviolet,
Visible, and Gravitational Astrophysics Program. M.M.N. acknowledges
support by the U.S. DOE.

\baselineskip=.30in


\end{document}